Theoretical Issues for Global Cumulative Treatment Analysis (GCTA)

5 August 2013

Jeff Shrager (shrager@stanford.edu)

Fellow, CommerceNet, Palo Alto, CA

Consulting Professor, Symbolic Systems Program

Stanford University, Stanford, CA

> "Every practicing physician conducts clinical trials daily as he is seeing patients."
> -- T.C. Chalmers, 1981

Introduction

Adaptive trials are now mainstream science (e.g,. Gallo, et al., 2006; Fiore, et al., 2011; Kim, et al., 2011; Berry, 2012; D'Avolio, et al. 2012), and more complex adaptive designs are being analyzed (e.g., Cai, et al, 2013), including ones that blur the distinction between classical phases (e.g., Berry, 2012). Although adaptive designs vary widely, their defining feature is that the particular treatment regimen (TR) that a patient is assigned to will vary depending upon the specific characteristics that describe that patient, combined with the running data regarding the performance of all arms. Just as in a classical randomized controlled trial (RCT) it is only statistically the case that a given patient will be placed into the TR that is most appropriate for them. However, in the adaptive case, this probability is updated continuously as data come in from all patients in every arm.

Recently, researchers have taken the adaptive trial concept to its natural conclusion, proposing what we will call "Global Cumulative Treatment Analysis" (GCTA) (e.g., Vickers and Scardino, 2009; Huber, 2013; Shrager and Tenenbaum, 2011; Tenenbaum and Shrager, 2011). In GCTA, just as in an adaptive trial, decision making and data collection and analysis are continuous and integrated; all available performance data for every possible TR is taken into account at every decision point for every patient. Similarly, like an adaptive trial, TRs are ranked in accord with the statistics of all this information, combined with what offers the most information gain, both for the patient at hand, and for every similar patient. Where GCTA differs from an adaptive trial, or, for that matter, from any trial design, is that *all patients are implicitly participants in the GCTA process, regardless of whether they are formally enrolled in a trial*.

Figure 1 summarizes the GCTA workflow. In the "normal treatment" (NT) scenario, whether or not involving a trial, a patient presents, some test results are obtained, and some choices are "computed", usually just based upon the treating physician's knowledge of the standard of care. These options are offered to the patient, who chooses among them, and treatment (or not) proceeds. Progress is monitored, and either the disease process resolves, or does not, and we cycle back to more tests, and so on. So far this is merely the typical way that medicine takes place. Also standard, although not as common, is the case where there are no standard treatment options, as is commonplace for cancer, and then the patient's options may include trials (whether RCTs or adaptive). So far we are still within the upper right-hand loop in Figure 1. Adaptive trials utilize a different method for computing choices than an RCT, specifically, in accord with the statistics of the ongoing trial, depicted by the box labeled "B". But otherwise, just as in an RCT, there is a clear distinction made between being in a trial or not (i.e., this is still within the "normal treatment", NT, model), and once in the trial, one is assigned to a particular TR and remains with that TR. This is, in fact, a stricter restriction that results from being enrolled in a trial than not being enrolled in one (i.e., NT): Trials, even adaptive ones, require that patients either remain on the assigned TR, or drop out of the trial. As we shall see, this is a very significant difference between trials of any sort, and the GCTA process.

Where the GCTA process differs from all the foregoing (i.e., from NT), is, first, all TRs are available to all patients at all times; although most of the time the statistics of experimental treatments for typical patients will be ranked very low. Secondly, unlike either an RCT or adaptive trial, but more like NT, decision making for each patient goes on continuously. However, unlike NT, this decision making takes into account all the available performance data over all TRs and all patients, including current ones. The other thing that makes the GCTA process unique is that, again, more like NT than like a trial, when there are no acceptable choices at all (circle "C"), one is free, indeed, encouraged, to undertake whatever analytical and treatment measures one can afford, and that data as well is fed back into the process. We represent this option ("C") by a circle with arrows to emphasize that what goes on here can be quite detailed dissection of the disease process (e.g., Blau and Liakopoulou, 2013), which can, within this cycle, involve complex omics, analysis, tumor boards, and so on. The issue from the present standpoint is not what specifically happens here, but that it is important that the data, what is decided, and why, are returned to the data stream so that near-future patients with similar characteristics can benefit from the findings and decision making inside this sub-analytical process.

Clearly the GCTA process relies upon a very carefully worked out choice ranking algorithm (box "A"), which we call the Adaptive Decision Algorithm (ADA). Below we analyze several critical issues surrounding this algorithm.

It is important to highlight the difference between the GCTA approach and a "big data/data mining" (BDDM) approach to treatment discovery and validation. In the GCTA approach, patients' choices are ranked at the point of care by a combination of the available information, combined with an estimate of the amount of information that can be obtained by ranking some choices over others, in careful accord with the available evidence. Consider, for example, the equipoise case where there is no information at all to distinguish between two TRs, that is, they are at 50/50 (or 1:1) with one another. In this case the BDDM approach expects there to be a random assortment of applications, and expects to be able to take advantage of the resulting data to demonstrate which TR performs better. In a perfectly efficient and transparent information economy, this might be the case, but we do not live in that world. The GCTA approach explicitly biases the TR rankings offered to a given patient to ensure that the appropriate level of experimental variability is injected into the distribution of patients with given characteristics. The GCTA approach expects that there will be surprises, but when such surprises happen – when a patient does not respond as expected (in either direction) – we can utilize the methods of precision medicine (or, more generally, just detailed analysis by whatever means) to dissect these cases in more detail. The BDDM approach has no expectations, and so there is no such thing as a surprise to it, except post hoc when it may be too late to go back and dissect the surprise, or, more importantly, change the course of treatment for that patient, if desired.

Numerous theoretical (i.e., statistical) issues arise in the design of a GCTA and the details of the ADA. There are, of course, numerous practical and ethical issues as well, but we focus here primarily on the theoretical issues. Whereas we know that these algorithms will be based upon Bayesian statistics, there remain numerous open questions as to the details of what should happen in certain common situations. I enumerate a number of such situations below, along with some preliminary thoughts on how they might be approached. Although I have tried to deal with these issues somewhat separately, they obviously interact with one another. In some cases I have pointed out this interaction and some of the implications, but in no case have I tried to describe the whole multi-dimensional problem, nor provide a covering solution. My goal here is to open this conversation, not to offer final solutions.

1. What is the right thing to recommend to patients who are in the middle of a complex treatment regimen (TR) when their TR is determined to be futile, or a different one is deemed superior? Similarly, when a new TR is admitted whose priors suggest that it might be superior to some of those already under analysis, under what circumstances should any patients be transitioned to this new treatment? What can we learn from what patients do in this case? (We assume, without repeating the phrase, that: "What can we learn from what patients do in this case?" is implicit in each of the following points.)

In the GCTA process patients (with advice from physicians) are making decisions about their TR on a continuous basis. Physicians, in turn are guided by the adaptive decision algorithm (ADA). When the ADA determines that a particular TR is futile, it is probably in the patients' best interest to transition them to a new, presumably superior TR. Once the possibility of treatment transition is accepted, there are numerous interesting implications for decision making and analysis. The TR to which they should move ought most likely be controlled by the same adaptive randomization algorithm used in preliminary intake, but, of course, informed by the new background that this patient has previously been under another TR; that is, the simplest thing to do is to treat this patient as through they are presenting for the fist time, although with newly expanded historical information. One theoretical question is: Should an adjustment be made to the forward-going statistical information generated by this patient to account for biases introduced by the patient having been already guided by the ADA previously? This interacts with the problem of complex treatment dynamics, esp. regarding sequential treatments, discussed in more detail below. In computing the transition choices for this patient, there may be evidence regarding specific changes in regimen, for example because a patient was not able to handle a particular treatment, and so there may be some evidence already regarding this specific transition. Regardless, such transitions offer us both additional data for the new TR to which they transition, as well as additional data regarding patients having been transitioned from the futile TR to new ones. It may be advisable to make such transitions in an incremental manner, for example, transitioning a handful of patients to a number of alternative TRs and observing their performance, and then resorting them as evidence accrues regarding these transitions.

2. What are the appropriate priors for new TRs or biomarker configurations (BCs) that come online?

Priors are, as usual, a central point of contention in Bayesian methods. There are several obvious simple ways to determine priors for new TRs (or BCs); for example, Spieghalter, et al. devote 40 page on the topic, and propose methods ranging from non-informative to expert-elicitation. The easy non-informative default is somewhat problematic because such priors are defined as non-informative with respect to some specific universe of information. For example, once there is some information regarding TRs that have been under analysis for some time, it is likely that the only reason that one would put forward the new TR is that the proponents have some prior reason to believe that it may be superior, or at least as good as, the current best contending TR. It might make sense, therefore, that the priors for the new TR be set to match the distribution of the current best contender, at least. This suggests that expert elicitation is possibly a better method than non-informative prior setting, although it, of course, has many issues of its own regarding expert bias. Certainly, as well, one would wish to utilize any available information from prior clinical experience with this TR, such as that available from toxicological, animal, or human dosing studies.

3. When a new BC is admitted, or an old one rejected, or a new marker deemed evidently related (or unrelated) to a specific TR, what do we recommend that patients with and without that BC, and in or out of the related TRs, do? What about patients whose BC overlaps, but is not exactly the same as a new or retired BC?

The introduction of a new biomarker, or, more generally, the discovery of a new relationship between a biomarker configuration (BC) and a treatment (hereafter called a BCTR: Biomarker Configuration Treatment Relationship), is a more complex case than the change in the state of belief of the efficacy of a TR. A given patient may fall under a specific BCTR or not, or a combination of BCTRs, or may have a BC that is similar to some degree (i.e., in some metric space) to the BC in a particular BCTR. Moreover, as a result of changes in the belief state with regard to various BCTRs, a particular patient may find him or herself (somewhat suddenly) undertaking the wrong TR, similar to the situation in question 1, above, but in the very much higher dimensional, and likely quite sparse and lumpy, space of molecular measures (more generally, patient characteristic vectors; PCVs). The simplest approach is, of course, an algorithm akin to nearest neighbor selection, but there are many other options (and parameters to set within each). Which is best for the expected distribution of PCVs will almost certainly require modeling to answer.

4. How should we deal with complex disease dynamics such as mutational escape recurrence in cancer? For example, a treatment that performs poorly in the short term might be better in the long run because it leads to less mutational escape and/or resistance to repeated or other treatments.

The successful operation of adaptive trials depends upon obtaining high quality outcomes (or proxy biomarkers for outcomes) as rapidly as possible. In a relatively slow disease, like most cancers, therefore, instead of waiting for "final" outcomes (usually something like 5 years disease free), proxy (or surrogate[1]) measures are almost always used, for example, tumor load or PSA status, which can be assessed right away through direct measurement, and moreover, tracked. However, unless the measure being used is a validated proxy – that is, something that has been shown to be highly correlated with the desired outcome – there is always the possibility of playing an early false positive. Moreover, because the ADA will tend to remove patients from *near term* underperforming TRs to *near term* better performing ones, at least according to the surrogate measures, the data that would offer the opportunity to discover a "late winner" is slowly (or sometimes rapidly!) depleted. We know that this could be a problem for some cancers (and possibly other disease), where recurrence has been observed to be more aggressive when the tumor is aggressively treated early on with only partly effective drugs (e.g., Nahta, et al, 2006).[2]

5. Closely related to question 4: How do we cope with the complexities and time dynamics of complex treatments that take place over months or years?

Treatment regimens (TRs) in most diseases, and especially in diseases such as cancer, often extend over weeks or months, and comprise multiple cycles of treatments of various sorts. Aside from there being innumerable parameters involved in such a treatment, most of which we hope are not very sensitively correlated with outcomes, the duration of such treatments complicates the data collection of proxy/surrogate measures, as well as the concept of decision point. On the one hand we might say that as much data should be captured as soon as possible, and every moment is a potential decision point, but there is certain to be variability, and most treatments have high short term variability, even if their long-term trend is toward overall improvement. Psychiatric drugs are especially problematic in this way because the measures of, say, depression, can vary so much from day to day that variability observed in a short term measurement window is almost certain to swamp possibly subtle long term effects. One way to deal with this is to admit aggregated, for example, windowed, measures instead of (or in addition to) single measures – measures, for example the patient's trend in reported depression. Unfortunately, this, again, slows down the measurement cycle, and introduces new parameters regarding the aggregation window.[3]

---

[1] We use "proxy" to refer to a short term measure (e.g., tumor load) that is validated to have a high correlation with respect to the desired long term measure (e.g., life expectancy). We use "surrogate" to refer to a short term measure (e.g., ECOG performance score) that is not so validated.

[2] Very interesting and applicable work in modeling disease evolution and combinatorial dynamic treatment is under intensive study in this area (e.g., Mumenthaler, et al., 2011).

6. How do we incorporate unordered categorical knowledge, such as peer review, expert opinion, patient opinions, pathway models, case reports, and so on, into the decision algorithms?

There is an abundance of important non-numerical knowledge in biology, such as peer review reports, expert opinion, patient opinions (i.e., experts in their own details), pathway models, case reports, and on and on. Sometimes there is embedded data in these that could be ferreted out, but unless this of a very specific sort, such as the statistical results of clinical trials, which is directly accessible through meta-analytic methods such as those employed by the Cochrane Collaboration (cf. Cochrane, 2013), the numerical content of these databases is not their primary output, and using that information out of context would likely lead to significant misinterpretations and errors. The available approaches to using this sort of knowledge generally rely upon some sort of domain specific "statisticalization" of the knowledge. (The Cochrane meta analysis method, for example, is a domain specific (re)statisticalization of published statistical results.) The most general form of statisticalization of categorical knowledge is to obtain survey scores from expert panels (e.g., Spiegelhalter, et al. 2003, ch. 5), and then use some aggregation of these scores as a proxy for the desired statistic. Another approach is to use counts. For example, Theibald, et al. (2003) scanned pubmed for co-occurrences between treatments, genes, and diseases, producing a set of "micro bayes nets", based upon relations from the PharmGKB knowledge base (Thorn, et al., 2013). This sort of count-based statisticalization leads naturally into the creation of graphical belief networks (e.g., bayes nets; Pearl, 2000, and many others). Such representations can represent very complex relationships, such as those found in signal pathways (e.g., Sachs, et al., 2002), and can be combined with molecular data such as expression data (e.g., Bay, et al. 2002). These graphical models can also be created through multiple-expert aggregation methods (e.g., Richardson and Domingos, 2003).

---

[3] S. Bagley (personal communication, 20130718) notes that a control theoretic method may be superior to a Bayesian one in certain aspects of this proposal, especially those involving time dynamics. Bayesian and control theoretic methods are closely related, but I admit to not understanding the mathematical subtleties well enough to be able to provide an informed analysis, beyond recognizing that this is a sensible suggestion. Working this out in detail will have to be left to experts in these methods.

7. How should we differentially weight case studies, small trials, large trials, individual patient experiences, and so on that may not have CIs?

We generally believe some sorts of evidence more than others; for example, we believe RCTs more than NEJM case reports, and NEJM case reports more than case reports in secondary journals, and case reports in any journal more than anecdotes, and anecdotes from doctors more than anecdotes from patients, etc. Meta analytical methodologies, such as the very well-worked out one utilized by The Cochrane Collaboration (2013) utilize the "GRADE" approach (Cochrane sec. 12.2.1), which provides 4 levels of evidence quality, ranging from "randomized trials; or double-upgraded observational studies", to "Triple-downgraded randomized trials; or downgraded observational studies; or case series/case reports." (Cochrane, sect. table 12.2.a). A somewhat more detailed approach was adopted by Mocellin, et al. (2010) who provide a specific "treatment ranking" algorithm based upon expert-encoded literature results, and utilizing a model weight on an exponential scale from meta-analysis down to *in vitro*. Rough attempts at statisticalization of more general evidence exists. For example, Mocellin, et al. (2010) asked experts to extract gene-drug-performance relationships from the melanoma literature, and then used a meta-analysis with 6 levels of rating, similar to the Cochrane GRADE system, to ranks drugs given a putative patient's biomarker configuration. Unfortunately, there is no good theory on how exactly to grade the more general range of knowledge and data that could and would be useful to incorporate in a real GCTA's ADA.

8. How are we to take account of choices made by doctors or patients that contradict the algorithmic guidance?

A very interesting problem occurs when a patient (with, or without the guidance of a physician) chooses to go against clear guidance, if any, of the ADA. There are three cases: First, they might not choose the TR that is ranked the highest by the ADA. Second, they might choose a TR that wasn't even in the list, but which is within the set of known TRs. Finally, third, they might choose to do something that the ADA has no data on at all. We will not deal in detail with any of these except to say that they will have to be dealt with in reality, because they occur. Also, we would like to have a way to get the patient's reason for choosing in this way back into the data stream in some manner, perhaps by offering a list of possible reasons that they could choose among, but there will have to be a "free text" option, and then all bets are off in open-ended analysis of this option. (For example, the patient could tell us that they know for a fact that the world is coming to an end tomorrow, and so they choose not to be treated. From the patient's point of view, this is perfectly reasonable, but how are we to integrate that into our algorithm? Perhaps get expert opinions on the likelihood and importance of this factor, and blend that – hopefully very small – factor back into the model?) In anycase, we note the importance of knowing that, and if possible why, a patient chooses to go against the ranking guidance provided by the algorithm. This leads naturally to the next topic:

9. How can the GCTA operating algorithms be validated at the same time as we are using them?

Because the ADA is a very complex algorithm, its correctness is critical. Unfortunately, as with any complex algorithm, we are unlike to be able to formally prove its validity (but see the extensive work on proving the correctness of complex algorithms by JS Moore, and colleagues, e.g., Kaufmann and Moore, 2004). Short of a proof, the two typical approaches are (a) simulation, and (b) measuring the effectiveness of the algorithm in the field (e.g., through a clinical trial).

A better approach is suggested by the previous issue, is to, in addition to monitoring outcomes which are likely to be very noisy, also monitor whether physicians and patients accept the algorithm's guidance, that is, choose among the highest ranked, essentially, equal, proposed TRs. This may be a more sensitive and less noisy measure than the outcomes themselves. Moreover, it ,maybe possible, and could be extremely useful, to gather information about why a physician or patient chooses to go against the guidance of the algorithm (although, as mentioned above, such reasoning, if delivered in free text, could be difficult make use of).

10. Decisions must include the availability and economics of treatments.

The unfortunate fact is that, regardless of the needs and desires of patients, physicians, and even pharmas, some treatments are hard to produce at high quality and/or quantity, some are expensive (for example, rare biologicals or surgical interventions), and many are both expensive and hard to produce in quality and/or quantity. This is especially true for early experimental biologics which, if they demonstrate promise, will become in high demand merely through the ADA process. But the cas where there is one TR that performs very well is of lesser concern than the case where there are a number of similarly performing TRs with widely varying costs, although the ADA will not reduce their demand to zero, it may be that the producers will decide to stop producing them because they are apparently not block-busters. Although this seems to be merely a practical matter, it has very real consequences for the viability of the GCTA model, and it's not clear whether the ADA should prefer to put more patients into the more expensive arm in order to ensure that there is adequate demand to keep producing it, or put less because in the long run, given two TR that are the same one prefers the less expensive one, although the whole point is that we do not know which is the better one in the long run at early junctures. The general point is that a correct ADA needs to take into account concerns that are not purely outcomes based, such as the economics of the TRs.

Conclusion

The Global Cumulative Treatment Analysis process is the completion of the thought of bringing clinical science into the internet age. It is, we contend not only the most efficient way of validating treatment options for future patients, but also the most ethical way to treat current patients. Whereas it may seem at first blush unethical to ask patients to participate in "experimental" treatments unless they are formally enrolled in a trial, one might instead ask why it is more ethical to keep patients, especially end stage patients, away from potentially helpful treatments just because those treatments have not been "fully" validated – if they are partly validated, give patients a partial chance to use them, and at the same time, add to our knowledge more rapidly than waiting for the end of a trial – even an adaptive one.

Of course, as with any new scientific process, there are numerous issues to be tackled, but these can, we believe, be worked out through discourse between statisticians, biologists, physicians, computer scientists, and regulators. This paper represents a preliminary contribution to that discourse.

Acknowledgements

Thanks to Steven Bagley, Tony Blau, Paul Howard, Peter Huber, Allan Schiffman, Joe Shrager, Marty Tenenbaum, and Andrew Vickers for their many useful comments.

Figures

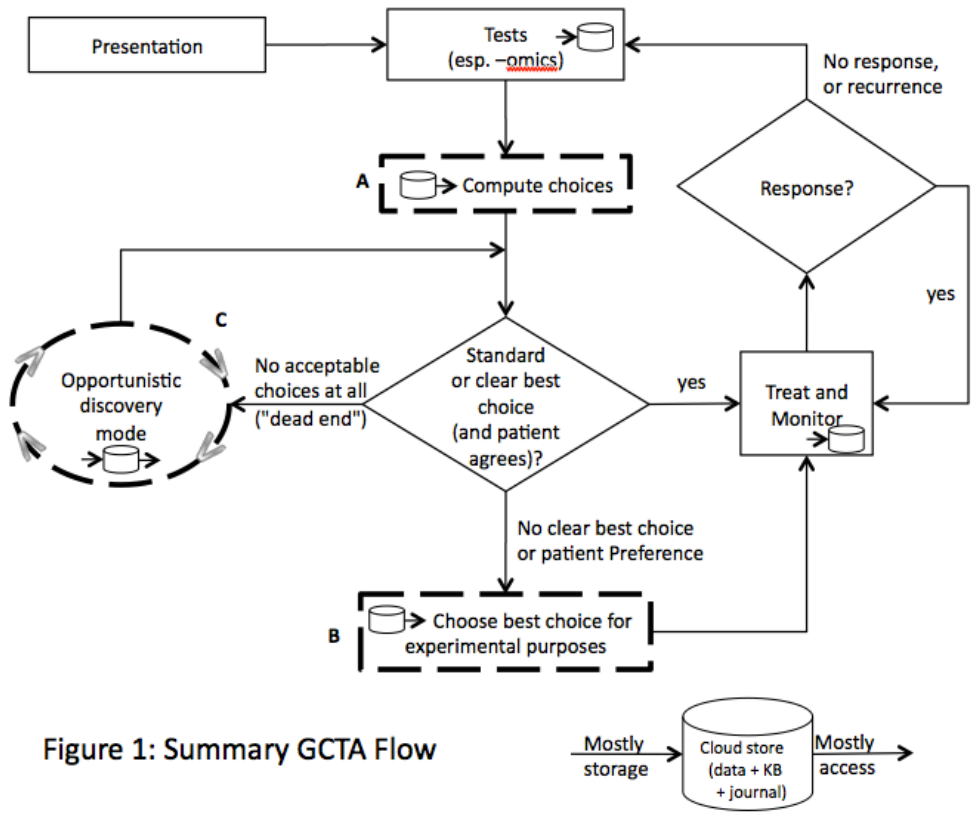

Figure 1: Summary GCTA flow. Typical treatment flow is presentation -> tests -> compute choices -> make choices -> treat/monitor -> response (or not), and back to more tests if necessary. GCTA flow (explained in more detail in the text) bases choice computation (A) on continuously collected data, and offers additional options when there is no preferred choice (B), and adds precision discovery mode for "dead ends" (C). Disk icons indicate where data is typically collected (left, inbound arrow), vs. primarily utilized (right, outbound arrow).